\documentclass[12pt, article]{aastex}
\shorttitle{IR transmission spectra for EGPs}
\shortauthors{Tinetti et al.}

\begin{document}

\title{Infrared Transmission Spectra for Extrasolar Giant Planets}

\author{Giovanna Tinetti\altaffilmark{1,2}
\email{tinetti@iap.edu} }
\author{Mao-Chang Liang\altaffilmark{3,4}
\email{mcl@gps.caltech.edu} }
\author{Alfred Vidal-Madjar \altaffilmark{2}
\email{alfred@iap.fr} }
\author{David Ehrenreich \altaffilmark{2}
\email{ehrenreich@iap.fr} }
\author{Alain Lecavelier des Etangs \altaffilmark{2}
\email{lecaveli@iap.fr} }
\and
\author{Yuk L. Yung\altaffilmark{3}}
\email{yly@gps.caltech.edu}
\affil{\altaffilmark{1} European Space Agency}
\affil{\altaffilmark{2} Institut d'Astrophysique de Paris,
CNRS (UMR 7095), Universit\'e Pierre et Marie Curie,  75014 Paris, France}
\affil{\altaffilmark{3} California Institute of Technology, 
Division of Geological and Planetary Sciences,
Pasadena, CA 91125, USA}
\affil{\altaffilmark{4} Research Center for Environmental Changes, Academia Sinica, Taipei 115, Taiwan}

\begin{abstract}
Among the hot Jupiters that transit their parent stars known to date, 
the two best candidates to be observed with transmission spectroscopy in the mid-infrared (MIR) are HD189733b and HD209458b,
 due to their combined characteristics of planetary density, orbital parameters and parent star distance and brightness.
Here we simulate transmission spectra of these two planets during their primary eclipse in the MIR, 
 and we present sensitivity studies
of the spectra to the changes of atmospheric thermal properties, molecular abundances and C/O ratios.
Our model predicts that the dominant species absorbing in the MIR on hot Jupiters are water vapor and carbon monoxide, 
and their relative abundances are 
determined by the C/O ratio. 
Since the temperature profile plays a secondary role in the transmission spectra of hot Jupiters compared to molecular abundances, 
future primary eclipse observations in the MIR of those objects might give an insight on EGP atmospheric  chemistry.
We find here that the absorption features caused by water vapor and carbon monoxide  in a cloud-free atmosphere,
 are deep enough to be observable by
the present and future generation of space-based observatories, such as Spitzer Space Telescope and 
James Webb Space Telescope. We discuss our results in  light of the capabilities of these telescopes. 
\end{abstract}

\keywords{ Atmospheric Effects,  Occultations, Radiative Transfer, Techniques: spectroscopic}

\section{Introduction}
Extrasolar Giant Planets (EGPs) are now being discovered at an accelerating pace \citep{schneider,butler}.
In particular,  an increasing interest has been focused on hot Jupiters that transit their parent stars, 
since they represent a valuable tool to determine key physical parameters of the EGPs, such as atmospheric composition and dynamics, thermal properties and presence of
condensates \citep{seager}.

The   most studied transiting  extrasolar planet, HD209458b, orbits a main sequence G-type star at 0.046 AU (period 3.52 days). It
 is the first one for which repeated transits across the stellar disk  were observed ($\sim$ 1.6\% absorption; \citet{henry,charbonneau}).
 Along with radial velocity measurements \citep{mazeh}, it was possible to determine 
  mass and radius ($M_{p}  \sim 0.69  \, M_{Jup}$, 
 $R_{p} \sim 1.4  \, R_{Jup}$), confirming the planet is a gas giant,  with one of the lowest densities discovered so far. Owing to this property, its very extended atmosphere is one of the best candidates to be probed with transit techniques. In particular the upper atmosphere extends beyond the Roche lobe, showing a population of escaping atoms. This discovery was possibly due to the
observed extraordinary deep absorptions in HI, OI and CII over the stellar emissions lines 
(15\%, 13\% and 7.5\% respectively; \citet{alfred1,alfred2}).
 The numerous follow-up observations of 
HD209458b, also include  the detection and upper limits
of absorption features in the deeper atmosphere \citep{charbonneaua, richardsona, richardsonb, demingb, richardsonc}. Most recently \citet{deminga}
detected the thermal emission of this planet with the Spitzer Space Telescope
during a secondary transit in the  24 $\mu$m band and \citet{richardsonc} detected the first primary
eclipse in the same band. 

To explain these observations several models
were proposed, including atmospheric photochemistry, thermal properties,  3-D circulation simulations, cloud and condensate height, escaping processes (e.g. \citet{fortney,burrows,liang1,lecav,yelle,tian,iro,seagerb}). For these reasons, the extrasolar planet HD209458b is the best known so far.
Additional observations are required, though, to constrain the past, present and future
 modeling effort.

The planet HD189733b, recently discovered by \citet{bouchy}, with mass $M_{p} \sim 1.15  \, M_{Jup}$
and $R_{p} \sim 1.26  \, R_{Jup}$,
orbits an early main sequence K star at 0.0313 AU. It is an exoplanet transiting the
brightest and closest star  discovered so far.

Here we focus our interest on both planets HD209458b and HD189733b, and we model the spectral absorption features in the Mid-Infrared (MIR) due to the most abundant atmospheric molecules
 during
their \emph{primary eclipse}, i.e., when the planet passes in front of  the parent star.
The use of transmission spectroscopy to  probe the upper layers of the transiting EGPs,
has been particularly successful in the UV and visible spectral ranges \citep{charbonneaua,richardsona,richardsonb,demingb,alfred1,alfred2}, and only very recently attempted in the MIR, in the 24 $\mu$m band, using Spitzer observations \citep{richardsonc}.
Two circumstances make it an approach worth considering. 
 First, the high surface temperatures, relatively small masses, and mostly hydrogen atmospheres of close-in EGPs imply large atmospheric scale heights. As a consequence, for spectral features that span  a reasonably wide bandwidth, so that the total
 photon flux is not too small,
this is a feasible and diagnostically powerful technique. 
Second, this is a complementary approach to the secondary eclipse observations.
 Transmission spectroscopy, taken during primary
eclipse, is sensitive to different parameters and regions of the atmosphere, compared to emission spectroscopy, on which the secondary eclipse method is based.

\paragraph{Oxygen versus Carbon $\to$ H$_2$O versus CO}
In a solar system like ours, a significant amount of water vapor (H$_2$O) can exist only in planetary atmospheres
at orbital distances less than 1 AU. The requirement is certainly met for the known transiting EGPs.
Carbon monoxide (CO) and methane (CH$_{4}$), and 
other photochemical products, such as carbon dioxide (CO$_{2}$) and acetylene (C$_{2}$H$_{2}$), are plausibly present in the atmospheres of EGPs, and possibly abundant
to be detected. These species have strong absorption bands in the MIR, and more importantly, in spectral regions
compatible with present and future space-based observations  such as the  Spitzer Space Telescope
or the James Webb Space Telescope \citep{jwst}.
Given O and C, H$_{2}$O and CO will be controlled mainly by the 
relative abundances of these two species.
\begin{itemize}
\item If C/O ratio is close to the solar, H$_{2}$O, CO and CH$_{4}$ abundances are determined by the thermodynamic equilibrium chemistry in the deep atmosphere 
 \citep{liang1,liang2}.
\item If C/O ratio is above solar, the atmospheric chemistry might change dramatically and, according to the scenario proposed by \citet{kuchner}, planets should show a significant paucity of water vapor in their atmospheres,  carbon rich species
should by contrast be enhanced. In particular, CO is expected to be the dominant carbon-bearing molecule at high temperatures and CH$_4$ the dominant at low temperatures \citep{kuchner}. 
\item If C/O ratio is below solar, the atmosphere is depopulated of carbon-bearing molecules and water-vapor 
is the dominant species between $\sim 10^{-10}$  to 1.5 bars.
\end{itemize}

\section{Description of the Model}
We have built a model of planetary atmosphere and  calculated the expected absorption of the stellar light when filtered through the planetary atmospheric
layers. This has been already discussed in the literature.
In particular for our simulations, we have used the geometry and the equations  described  in \citet{brown} (fig. 1, configuration 2) and \citet{david} (sec. 2.1, fig. 1). Our cloud and haze-free atmospheres were divided in forty layers spanning from $\sim 10^{-10}$  to 1 bars.

 Photochemical models are used to determine the molecular abundances of 33 species above $\sim$1 bar altitude level. We start with four parent molecules H$_2$, CO, H$_2$O, and CH$_4$. Their relative abundances are determined by thermochemistry in the deep atmosphere, and are fixed as our lower boundary condition. Chemical reactions and eddy mixing profiles are taken from \citet{liang1,liang2}. Details of the model can be referred to \citet{liang1,liang2} and references contained therein. For the simulation of the photochemistry of HD209458b, we adopt the solar spectrum. For HD189733b we use the spectrum of HD22049,
 which is a K2V star similar to  HD189733 \citep{segura}.
 We have repeated our calculations for three temperature-pressure profiles
(fig. \ref{model} right), to  test the sensitivity of our results to these assumptions. The modeled chemical abundances  show a negligible dependence on temperature (see
\citep{liang2})

The absorption coefficients in the MIR were estimated  using
  a line-by-line model, LBLABC \citep{meadows},  that generates
 monochromatic gas absorption coefficients  from molecular line lists -HITEMP, \citet{rothman}-, 
for each of the gases present in the atmosphere.  
\section{Results}
\begin{figure}
    \centering
\mbox{    \includegraphics[width=9 cm]{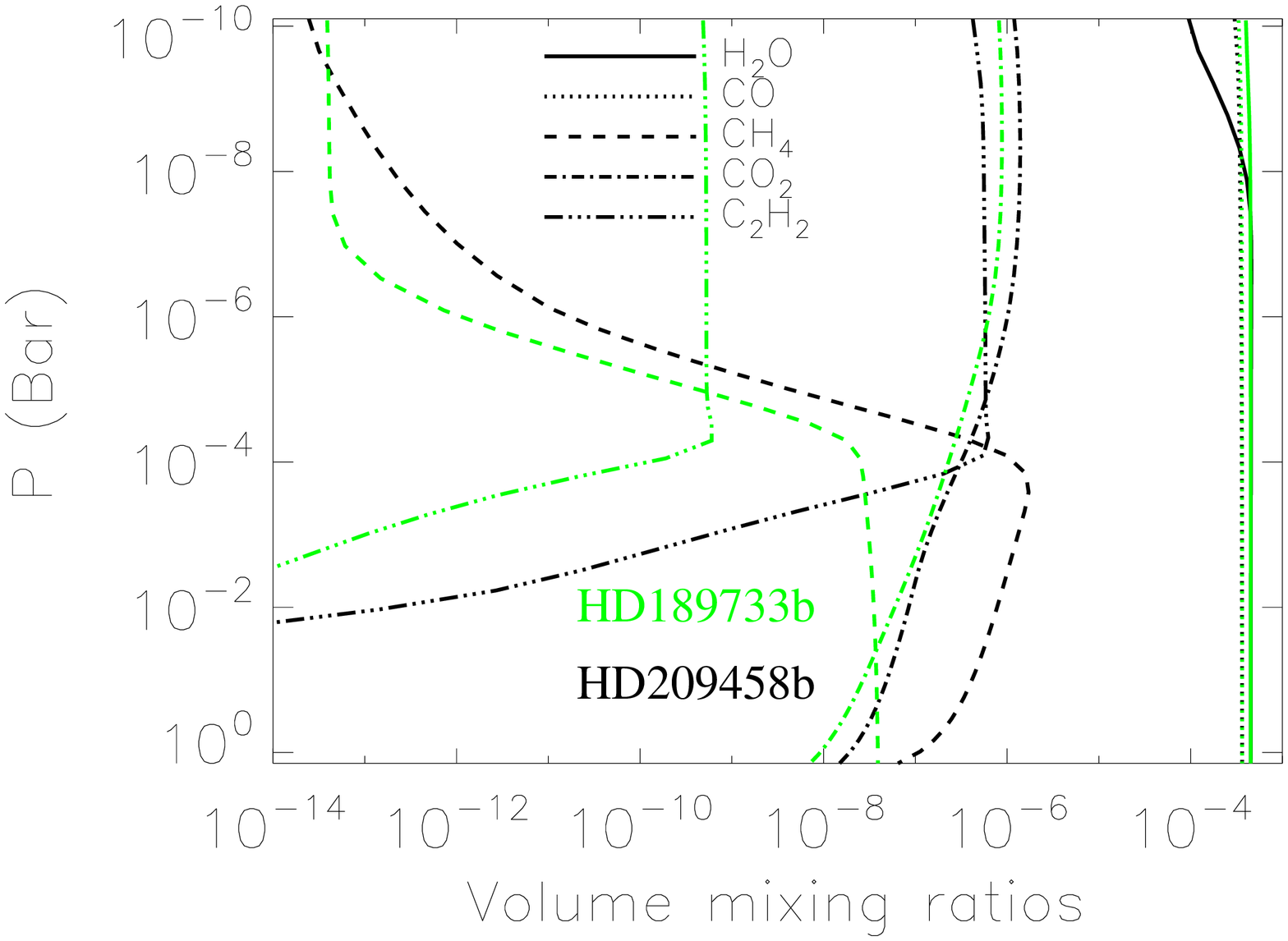} 
 \includegraphics[width=6.5cm, angle=90]{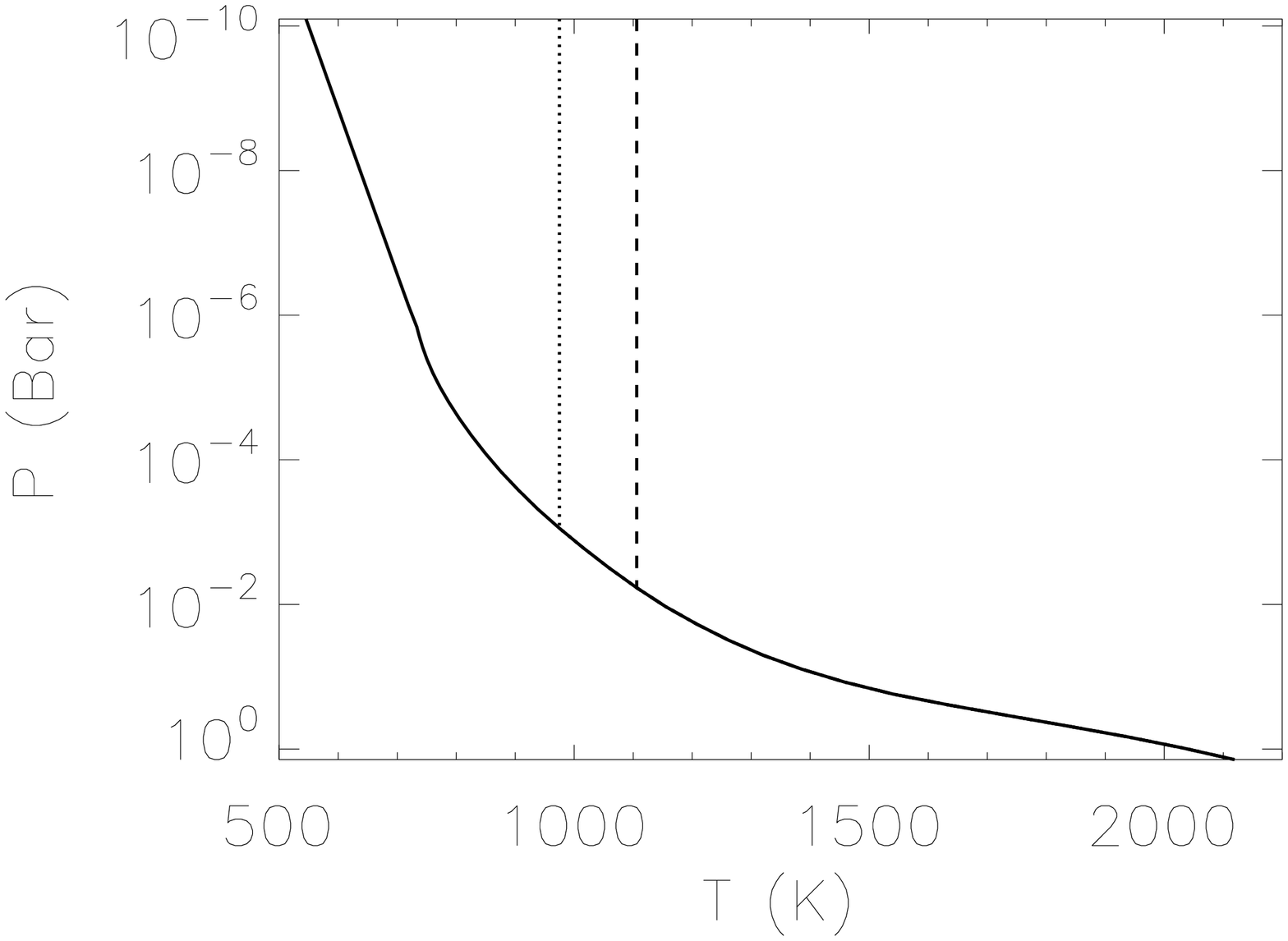} }
\caption{{\footnotesize  Left: Profiles of H$_2$O (solid line), CO (dotted line), CO$_2$ (dot-dashed line), CH$_4$ (dashed line), and C$_2$H$_2$
(triple-dot-dashed line) for planets HD209458b (black lines) and HD189733b  calculated with
the photochemistry model described in \citet{liang1,liang2}.
Right: temperature-pressure profiles used for our simulations. Solid line: standard, dotted line: hot profile,
dashed line: very hot profile.} } \label{model}
 \end{figure}

Fig. \ref{model} (left) shows the molecular profiles of H$_{2}$O, CO, CH$_{4}$, CO$_{2}$ and C$_{2}$H$_{2}$ for both  planets HD189733b and HD209458b, 
calculated by the photochemistry model with solar
C/O ratio as boundary condition. On HD189733b, H$_{2}$O,  CH$_{4}$ and C$_{2}$H$_{2}$ are more abundant in the upper atmosphere compared to HD209458b,
 since HD189733 is a later type star, therefore the photo-dissociation processes occurring in the  atmosphere of that planet are less significant.

\begin{figure}[h!]
    \centering
\mbox{  \includegraphics[width=8.5cm]{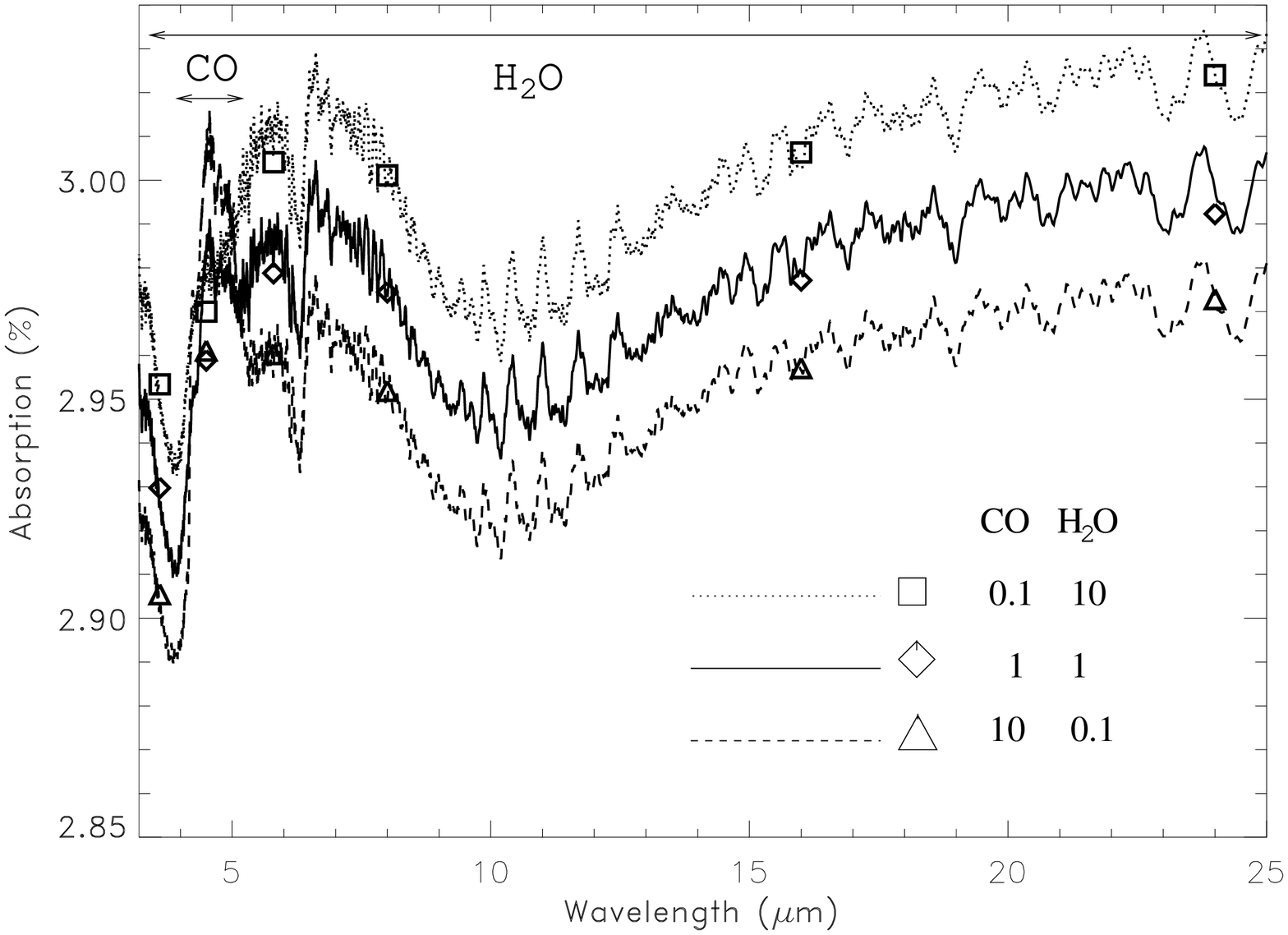}
\includegraphics[width=8.5cm]{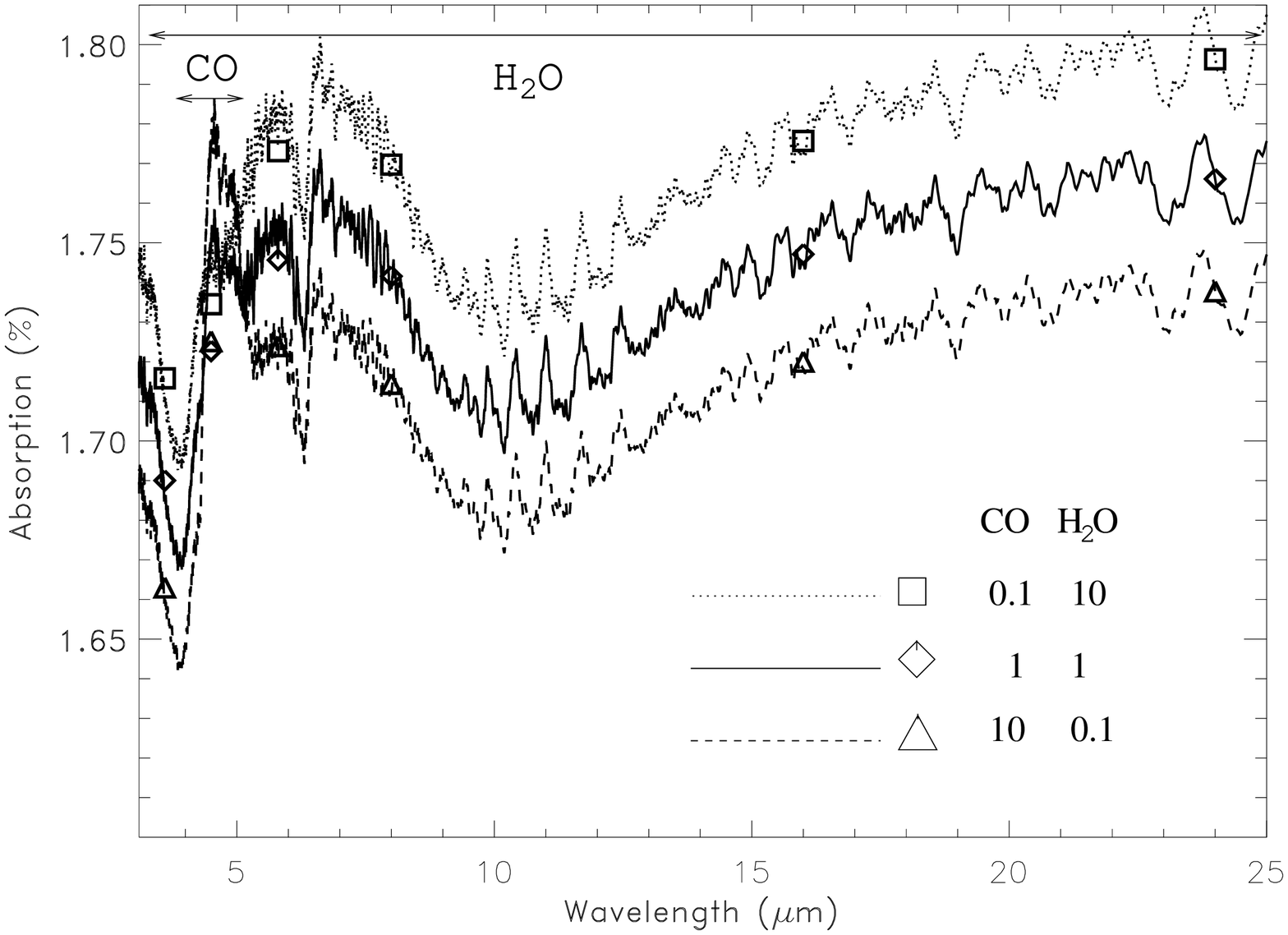} }
\caption{{\footnotesize 
Modeled spectral absorptions of water vapor and CO in the atmospheres of HD189733b (left) and
HD209458b (right) during their transits. Solid line: standard case, C/O ratio = solar (fig. \ref{model} left); dashed line: CO is over- and H$_2$O under-abundant  of a factor 10;
dotted line: the opposite.
 White triangles, squares and rhombi indicate the spectral absorptions averaged over the
Spitzer's IRAC, IRS and MIPS bandpasses, centered at 3.6, 4.5, 5.8, 8, 16 and 24 $\mu$m.  
Only in the 4.5 $\mu$m band,
where CO shows a strong absorption, the triangle and rhombus overlap.
} } \label{ratio}
 \end{figure}

\emph{Sensitivity to molecular abundances and C/O ratio.} Figs.~\ref{ratio}  show the predicted absorption signatures due to water vapor and CO on the planets HD189733b and HD209458b. The three plots
compare the spectral absorptions of these two species when C/O ratio is solar (standard case, solid line, see fig. \ref{model} left for mixing ratios)
and when is below  and above  solar. As specific examples, we have assumed  H$_2$O to be 10 times more  and at the same time CO 10 times less abundant than the standard case (dotted line), 
and vice-versa (dashed line). When we increase/diminish H$_2$O and CO of a factor 10, 
 the  absorption is  increased/decreased by a constant $\sim$ 0.03~\% through all the selected wavelength range. 

In these figures, we show also the signatures (white rhombi, squares and triangles)
 relative to the three cases (standard, C/O ratio below and above solar) averaged over the IRAC, IRS and MIPS bandpasses
 (centered at 3.6, 4.5, 5.8, 8, 16 and 24 $\mu$m),  the instruments
 on board the Spitzer Space Telescope (see table 1 for the calculated absorptions). Water vapor has strong absorption lines through all the selected spectral range in the MIR, 
the CO signature appears only in a narrower wavelength interval, where the IRAC 4.5 $\mu$m bandpass  is centered.
 When the C/O ratio is above solar,
the triangle is expected to appear above the rhombus in that band, indicating the strong CO contribution to the total absorption (table 1, numbers in bold).

\begin{figure}[h!]
    \centering
\mbox{  \includegraphics[width=6.3cm, angle=90]{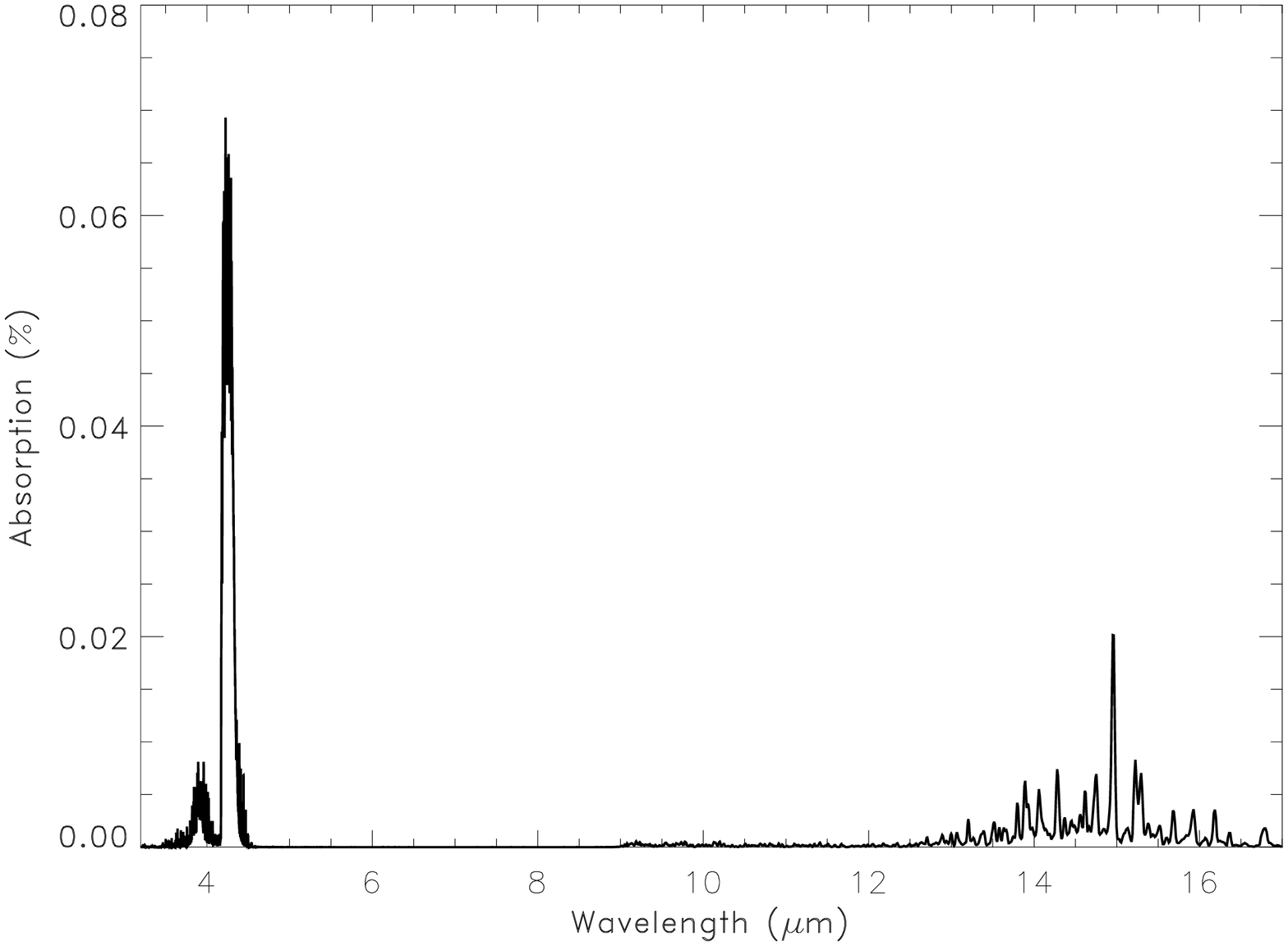}
\includegraphics[width=6.3cm, angle=90]{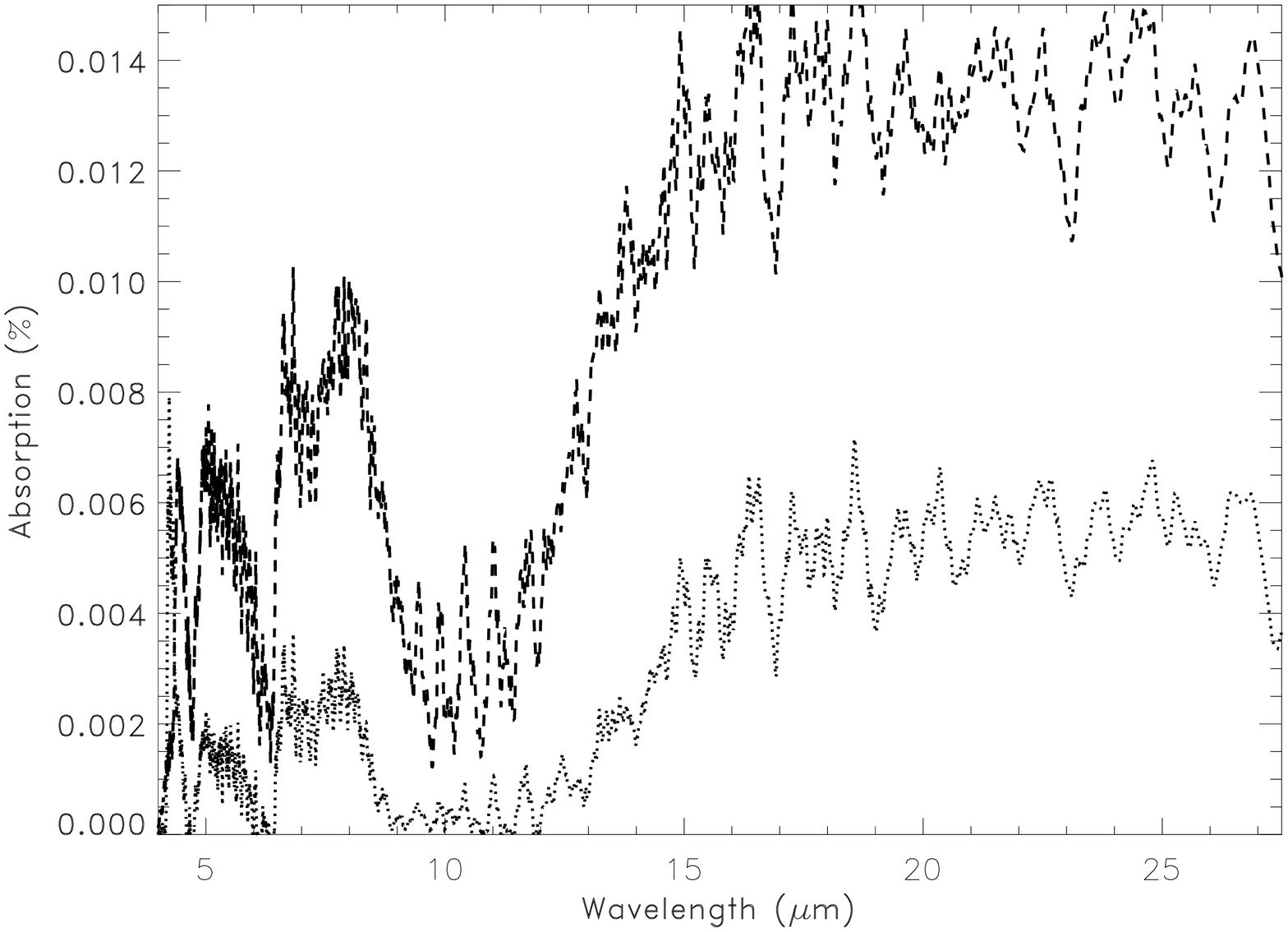} }
\caption{{\footnotesize 
Left: residue spectrum from difference of 2 curves:  dashed line plot in fig. \ref{ratio} top  with and without the contribution of
 CO$_2$.  CO$_2$ is assumed  to be, like CO,  over-abundant  by a factor 10. Right: residue spectra from difference of 2 curves. Dotted line: difference from hot and standard profiles;
dashed line: difference from very hot and standard profiles
 (fig \ref{model} right). The discrepancies amidst the curves is negligible compared
to the changes in molecular abundances (fig. \ref{ratio}).
} } \label{co2}
 \end{figure}

Note that we included the species  CH$_{4}$, CO$_{2}$ and C$_{2}$H$_{2}$ in our calculations, 
which strongly absorb in the MIR.
However their abundances are  too small compared to CO and H$_2$O, and they are masked by these species. For  example, we show in fig. \ref{co2} (left)
the contribution due to CO$_2$  in the case of C/O above solar ratio. CO$_2$ is increased by a factor 10 here, to follow CO behavior 
consistently with chemistry predictions \citep{liang1}. When CO$_2$ is present, an increase in absorption of $\sim$ 0.02~\% is found  in the band centered on
15 $\mu$m and a very narrow peak reaching 0.07~\% is visible at shorter wavelengths. Although not negligible, the contribution due to CO$_2$ is masked by CO and water.

\emph{Sensitivity to temperature.} The effects due to temperature variations (fig. \ref{co2} right) are negligible compared to the changes in molecular mixing ratios. Temperature  plays a secondary role
in the determination of the optical depth: it affects the absorption coefficients and the atmospheric scale heights (see eq. 2 in \citet{david}).  
For HD189733b, the discrepancy between the standard and the hot profile is less than 0.005\%, and between the standard and the very hot profile
 has a maximum of 0.014\% in the  15-30 $\mu$m range (fig. \ref{co2} right). 
Analogous results are obtained for HD209458b.

\section{Discussion}
In our model we did not include the contribution of hazes or clouds \citep{ackerman,lunine,fortneya}.  
Due to their presence, the atmospheric optical depth might increase, partially masking the absorption features due to atmospheric molecules.
In the case of water vapor and CO, only clouds/hazes lying at  altitudes higher than 1 bar might affect our results. Predictions of cloud/haze is
particularly difficult for EGPs, since the few observations we have are not sufficient to constrain all the cloud microphysics and aerosol parameters. 
Moreover, the planetary limb observable during the star occultation might show the signatures of both the night and day side of those planets, which are presumably tidally locked \citep{iro}. The thermal profiles, hence the condensate dynamics, might be very different on the two sides. Consequently, a more complete model able to predict
 cloud  and haze location and optical characteristics,
 should contain a 3-D dynamical simulation of the atmosphere. In this paper we  limit our simulations to the cloud-haze free atmosphere,
with the caveat they might be perturbed by  the possible presence 
(constant or variable) of optically thick particles in the atmosphere above 1 bar pressure.

Same considerations are valid for the thermal profiles. A extensive literature is available on $T-P$ profiles for EGPs  at pressures from $\sim$1 bar to
 10$^{-4}$-10$^{-6}$ bars, 
most recently
including 3D dynamical effects \citep{showman,iro,cho,burrowsa}. For transmission spectroscopy in the MIR, we need to consider also  the contribution of the 
upper atmosphere. The $T-P$ profiles calculated by \citet{tian,yelle} suggest that the  trend for the atmospheric temperature is to increase in the exosphere.
For our simulations, we use a $T-P$ profile compatible with the lower atmosphere models cited above up to 10$^{-3}$-10$^{-4}$ bars, and then we consider three cases:
the atmospheric temperature decreases up to 10$^{-10}$ bars (standard), the atmosphere is isothermal (hot, very hot profiles). 
Our results, show that the  differences among the spectra
calculated with  the three profiles are within 0.009~\% for $\lambda \le 14 \mu$m and within  0.014~\% at longer wavelengths,
 so we are confident our simulations will not significantly change using a  more refined 
thermal structure.

Our model atmospheres extend to 10$^{-10}$ bars, where non local thermodynamic equilibrium (non-LTE) effects might occur \citep{kutepov}. However, if we truncate our calculations to  10$^{-5}$ bars, we obtain a maximum
error of $\sim 0.02 \%$  at 30  $\mu$m (no discrepancy  for wavelengths shorter than 20 $\mu$m), indicating that our calculated absorptions in LTE regime are correct at first order approximation. 

In order to detect the presence of water vapor and CO on HD189733b and HD209458b in the MIR, an extra absorption of $\sim$ 0.15~\% is expected to be added to the 
2.85~\% and 1.6~\% due to the optically thick  disks at 1 bar atmospheric level.
 To estimate the chemical abundances, an accuracy of at least 0.03~\%  is  needed.
By inspection of the relative
absorption of the IRAC 4.5 $\mu$m bandpass with respect to the others, we might be able to infer the C/O ratio, but in this case 
 an extremely high S/N is required.
 HD189733 is a bright K0V star of magnitude K = 5.5. We estimate the brightness in the four IRAC bands to be of the order of 1850, 1100, 730 and 400 mJy at
3.6, 4.5, 5.8 and 8  $\mu$m respectively. For HD209458, a G0V star with K-magnitude of 6.3,  the IRAC predicted fluxes are 878, 556, 351 and 189 mJy.
According to these numbers, a better S/N should be obtainable for HD189733b \citep{demingc}, and this makes   HD189733b a better candidate for observations. 

\section{Conclusions}
In this paper we have presented simulations of transmission spectra of two extrasolar giant planets during their transit in front of their parent star.
According to our calculations, we estimate an excess absorption in the 
IR of up to 0.15 \%
for HD189733b  and up to 0.12 \% for HD209458b (C/O ratio $\sim$ solar), in addition to the 
nominal 2.85 \% and 1.6 \%
absorptions measured at shorter wavelengths.
 If water were far less abundant, 
 other species
 might be observable, depending on their mixing ratios. Among them, CO$_{2}$,  CH$_{4}$ and C$_{2}$H$_{2}$ are the best candidates. 

According to our simulations, transmission spectra of EGPs in the MIR are    sensitive
to molecular abundances and less  to temperature variations. Temperature influences the transmission spectrum  by
way of its influence on the atmospheric scale height, as discussed by Brown  (2001), and on the absorption coefficients. 

If water vapor and CO are as abundant as  photochemical models predict,
we expect they can be detected with the IRAC, IRS and MIPS instruments on board the Spitzer Telescope and with future telescopes like JWST.
Moreover, if an accuracy of 0.03~\% is obtainable, future observations  may give  a first direct estimate of  H$_{2}$O and CO abundances in the upper atmosphere of EGPs and possibly -depending on their 
mixing ratios- a constraint on  CO$_{2}$, CH$_{4}$ and C$_{2}$H$_{2}$.

\acknowledgments
\section*{Acknowledgments}
We would like to thank the anonymous referee for his help to improve the paper,
L. S. Rothman for having provided the HITEMP data list,
R. Ferlet, J. M. D\'esert, F. Bouchy, G. Hebrard, A. Noriega Crespo and S. Carey, for their valuable inputs, and C. D. Parkinson for useful comments.
G. Tinetti is currently sponsored by the European Space Agency.
M. C. Liang and Y. L. Yung  are supported by the NASA grant NASA5-13296 to  the California Institute of Technology.

\clearpage

\begin{deluxetable}{lrrrrrrrr}
\tablewidth{40pc}
\tablecaption{Calculated absorptions averaged over IRAC, IRS \& MIPS bandpasses.We recall that 2.85~\% and 1.6~\% are the nominal absorptions
 due to the optically thick  disks at 1 bar atmospheric level.
 In bold the absorptions when C/O ratio is above solar and CO strongly contributes in the 4.5 $\mu$m IRAC band.}
\tablehead{
\colhead{ } & \colhead{C/O ratio}               & \colhead{3.6 $\mu$m} &
\colhead{4.5 $\mu$m} & \colhead{5.8 $\mu$m}    &   \colhead{8 $\mu$m}    & \colhead{16 $\mu$m}    &
\colhead{24 $\mu$m}}

\startdata
HD189733b & $<$ solar & $2.953  $ & $2.970 $ & $3.004 $ & $3.001 $ & $3.006  $  & $3.024 $ \\
HD189733b & solar & $ 2.930 $ & $2.959 $ & $2.978 $ & $2.974  $ & $ 2.977 $ & $2.992 $ \\
HD189733b & $>$ solar & $ 2.905 $ & $\mathbf{2.961}$ & $2.960$ & $ 2.952$ & $2.957$  & $ 2.973$ \\
\\
HD209458b & $<$ solar & $1.715$ &  $1.734$ & $1.773$ & $1.770$ & $1.776$ & $ 1.796 $ \\
HD209458b & solar & $1.690$ & $1.722$ & $1.746$ & $1.741$ & $ 1.747 $ & $ 1.766 $ \\
HD209458b & $>$ solar & $ 1.663 $ & $\mathbf{1.725}$ & $1.724$ & $1.714$ & $1.720$ & $1.738$ \\
\enddata
\end{deluxetable}

\begin{thebibliography}{}
\bibitem[Ackerman and Marley(2001)]{ackerman} Ackerman A.S. and Marley M. S. (2001), \apj, 556, 872.  
\bibitem[Bouchy et al.(2005)]{bouchy} Bouchy  F., Udry S., Mayor M., Moutou C., et al. (2005),  A\&A,  444, L15.
\bibitem[Brown(2001)]{brown} Brown T. M. (2001), \apj, 553, p 1006.
\bibitem[Burrows et al.(2003)]{burrows} Burrows A., Sudarsky, D., and Hubbard, W. B., (2003), \apj.,  594, 545.
\bibitem[Burrows et al.(2006)]{burrowsa} Burrows A., Sudarsky D. and Hubeny I.  (2006), \apj.,  accepted, astroph 0607014.
\bibitem[Butler et al.(2006)]{butler} Butler R. P., J. T. Wright, G. W. Marcy, et al. (2006), \apj, 646,  505.
\bibitem[Charbonneau et al.(2000)]{charbonneau} Charbonneau D., Brown T. M.,Latham D. W. and Mayor M. (2000), \apj, 529, L45.
\bibitem[Charbonneau  et al.(2002)]{charbonneaua} Charbonneau D., Brown T. M., Noyes R. W.  and Gilliland R. L. (2002),  ApJ 568, 377.
\bibitem[Cho et al.(2006)]{cho} Cho J. Y. K., Menou K., Hansen B. M. S. and Seager S. (2006),  astroph 0607338.
\bibitem[Deming et al.(2005a)]{deminga} Deming D., Seager S., Richardson L. J. and Harrington J.,  (2005a), Nature, 434, 740.  
\bibitem[Deming et al.(2005b)]{demingb} Deming D., Brown T. M., Charbonneau D. et al. (2005b), \apj, 622, 1149.
\bibitem[Deming et al.(2006)]{demingc} Deming D., Harrington J., Seager S. and Richardson L. J.,  \apj, 644, 560-564
\bibitem[Ehrenreich et al.(2006)]{david} Ehrenreich D., Tinetti G., Lecavelier A. et al. (2006), A\&A, 448, 379.
\bibitem[Fortney et al.(2003)]{fortney} Fortney J. J., Sudarsky D., Hubeny I, et al. (2003), \apj, 589, 615.
\bibitem[Fortney et al.(2005)]{fortneya} Fortney  J. J., Marley M. S., Lodders K., Saumon D. and Freedman R.  (2005), \apj, 627, L69.
\bibitem[Gardner et al.(2006)]{jwst} Gardner J. P., Mather J. C.; Clampin M. et al., Space Science Reviews, in press.
\bibitem[Henry et al.(2000)]{henry} Henry G. W., Marcy G. W., Butler R. P. and Vogt S.S. (2000), \apj, 529, L41.
\bibitem[Iro et al.(2005)]{iro} Iro N.,  B\'ezard B.,  Guillot T. (2005), A\&A,  436,   719.  
\bibitem[Kuchner and Seager(2006)]{kuchner} Kuchner M. and Seager S. (2006),  astroph 0504214, \apj, submitted.
 \bibitem[Kutepov et al.(1998)]{kutepov} Kutepov A. A., Gusev O. A., and Ogibalov V. P., 1998, J. Quant. Spectrosc. Radiativ. Transfer, 60, 199-220. 
\bibitem[Lecavelier et al.(2004)]{lecav} Lecavelier des Etangs A., Vidal-Madjar A., McConnell J. C. and H\'ebrard G., et al. (2004), A\&A,  418, L1.
\bibitem[Liang et al.(2003)]{liang1} Liang M. C., Parkinson C. D., Lee A. Y.-T. et al. (2003), \apj, 596, L247. 
\bibitem[Liang et al.(2004)]{liang2} Liang M. C., Seager S., Parkinson C. D., et al (2004), \apj, 605, L61. 
\bibitem[Lunine  et al.(1989)]{lunine} Lunine J. I., Hubbard W. B., Burrows A., et al. (1989), \apj, 338, 314. 
\bibitem[Mazeh et al.(2000)]{mazeh} Mazeh T., Naef D., Torres G., et al. (2000), \apj, 532, L55.  
\bibitem[Meadows and Crisp(1996)]{meadows}  Meadows V. S.  and  Crisp D. (1996),
JGR,  101, 4595.
\bibitem[Richardson et al.(2003a)]{richardsona} Richardson L. J., Deming D. and Seager S., (2003a),  \apj, 597, 581-589.  
\bibitem[Richardson et al.(2003b)]{richardsonb} Richardson L. J., Deming D.,  Wiedemann G., et al. (2003b), \apj, 584, 1053.   
\bibitem[Richardson et al.(2006)]{richardsonc} Richardson L. J., Harrington J., Seager S. and Deming D.,  (2006), \apj, 649, 1043-1047. 
\bibitem[Rothman et al.(in preparation)]{rothman} Rothman L. S., C. Camy-Peyret, J.-M. Flaud, et al., HITEMP, the High-Temperature Molecular Spectroscopic 
Database is being prepared for the Journal of 
Quantitative Spectroscopy and Radiative Transfer.
\bibitem[Schneider(2006)]{schneider} Schneider J. (2006), The Extrasolar Planets Encyclopaedia, http://exoplanet.eu/
\bibitem[Seager and Sasselov(2000)]{seager} Seager S. and Sasselov D. D. (2000), \apj, 537, 916.
\bibitem[Seager et al.(2005)]{seagerb} Seager S., L. J. Richardson , B. M. S. Hansen et al. (2005), \apj, 632, 1122.
\bibitem[Segura et al.(2003)]{segura} Segura A.,  Krelove K., Kasting J. F.,  et al. (2003), Astrobiology, 3(4),  689 -708 
\bibitem[Showman and Guillot(2002)]{showman} Showman, A. P. and Guillot T., (2002), A\&A, 385, 166.
\bibitem[Tian et al.(2005)]{tian} Tian F., Toon O. B., Pavlov A. A. and De Sterck H. (2005), \apj, 621, 1060.
\bibitem[Vidal-Madjar et al.(2003)]{alfred1} Vidal-Madjar A., Lecavelier des Etangs A., D\'esert J. M., et al. (2003), Nature 422, 143.
\bibitem[Vidal-Madjar et al.(2004)]{alfred2} Vidal-Madjar A.,  D\'esert J.-M.; Lecavelier des Etangs A. et al.(2004),  \apj, 604, L69.
\bibitem[Yelle(2004)]{yelle} Yelle R. (2004), Icarus 170, 167.
\end{thebibliography}
\end{document}